# Spin-exchange collisions of submerged shell atoms below 1 Kelvin

J. G. E. Harris,[1] S. V. Nguyen,[2] S. C. Doret,[2] W. Ketterle,[3,4] and J. M. Doyle[2,4]

[1] *Departments of Physics and Applied Physics, Yale University, New Haven, CT 06520*

[2] *Department of Physics, Harvard University, Cambridge, MA 02138*

[3] *Department of Physics, MIT, Cambridge, MA 02139*

[4] *Harvard / MIT Center for Ultracold Atoms, Cambridge, MA 02138*

Angular momentum changing collisions can be suppressed in atoms whose valence electrons are submerged beneath filled shells of higher principle quantum number. To determine whether spin-exchange collisions are suppressed in these "submerged shell" atoms, we measured spin-exchange collisions of six hyperfine states of Mn at temperatures below 1 K. Although the 3d valence electrons in Mn are submerged beneath a filled 4s orbital, we find that the spin exchange rate coefficients are similar to those of Na and H (which are non-submerged shell atoms).

PACS Numbers: 39.25.+k, 34.50.Pi, 34.10.+x, 31.70.-f

Quantum degenerate atomic gases are most commonly realized by evaporatively cooling atoms in a magnetic trap. Efficient evaporative cooling requires that angular momentum changing collisions (which can be driven by dipolar, second-order spin-orbit, spin-exchange, and anisotropic electrostatic interactions[1,2,3,4]) occur at much lower rates than elastic collisions. This requirement is usually met by trapping an atomic species with a small magnetic moment (to minimize dipolar interactions) and a ground state with orbital angular momentum $L = 0$ (which removes anisotropic electrostatic interactions to lowest order). These atoms are then trapped in "stretched" hyperfine states which are immune to spin-exchange relaxation. Alkali atoms meet these criteria, and are by far the most widely-studied atoms in the ultracold regime.

Achieving quantum degeneracy in a wider range of atomic species would enable the study of new types of quantum fluids,[5] quantum computing architectures,[6] and searches for time dependence of fundamental constants.[7] As a step towards this goal recent work[8,9,10] showed that some angular momentum changing collisions are suppressed for "submerged shell" atoms - species whose valence electrons lie at smaller radii than filled shells of higher principal quantum number. In collisions between He and a "submerged shell" atom with $L \neq 0$, the filled outer shells shield the anisotropic electrostatic interaction, reducing the rate of angular momentum changing collisions by $10^4 - 10^6$. As result, $L \neq 0$ submerged shell atoms could be cooled via elastic collisions with a He buffer gas while maintaining the orientation of their magnetic moment. Atomic Tm, Er, Nd, Tb, Pr, Ho, and Dy (4f valences shielded by filled 5s and 6s shells) were all magnetically trapped for the first time using this approach. Collisions between He and Ti

(3d valence shielded by filled 4s shell) showed a comparable suppression of the anisotropic electrostatic interaction.[8,9]

It is an open question whether whether a submerged valence suppresses other types of angular momentum changing collisions, such as spin-exchange. Another open question is whether a submerged valence suppresses angular momentum changing collisions of *pairs* of submerged shell atoms (as opposed to collisions between He and a submerged shell atom, as in Refs. 8 - 10). These questions are central to determining whether submerged shell effects can extend the range of atomic species and hyperfine states which can be efficiently evaporatively cooled. They are interesting from the point of view of atomic collisions generally.

To address these questions we have trapped several hyperfine states of atomic Mn, a submerged shell atom with a half-filled 3d valence within a filled 4s shell.[11] We determined spin-exchange and dipolar rate coefficients by measuring trap losses. Unlike previous studies of collisions involving submerged shells,[8,9,10] Mn has $L = 0$, leading to isotropic electrostatic interactions. This allows us to focus more cleanly on spin-exchange. Also unlike previous studies, we measure the properties of collisions *between* submerged shell atoms, as opposed to collisions between submerged shell atoms and He. By performing these measurements at relatively high temperatures (~ 1 K) where many partial waves contribute to the collisions, our results are more likely to reflect generic atomic structure effects than measurements in the ultracold regime which may be sensitive to "accidental" cancellations of a single partial wave.[12]

The half-filled 3d valence of Mn gives a $^6S_{5/2}$ ground state. The $S = 5/2$ electronic spin is coupled via hyperfine to the $I = 5/2$ nuclear spin (Mn has only one stable isotope).

Fig. 1(a) shows a calculation of the magnetic field dependence of the 36 states in the ground electronic manifold. This calculation includes Zeeman, magnetic dipole hyperfine, and electric quadrupole hyperfine terms.[13] In our experiment the overwhelming majority of the atoms are in *B* fields where the Zeeman energy is much greater than the hyperfine splitting, so we label the eigenstates $|1\rangle$ to $|36\rangle$ starting from the highest-energy state in the large-*B* limit (Fig. 1(a)).

Atomic Mn was cooled and trapped using a $^3$He buffer gas apparatus described extensively elsewhere (Fig. 1(b)).[12,13,14,15,16] Once loaded into the anti-Helmholtz magnetic trap, the lifetime of trapped Mn atoms in their most low-field seeking states (i.e., those corresponding to $m_S = 5/2$) was not limited by collisions with background $^3$He. However the density of $^3$He was kept sufficiently high so that Mn - $^3$He collisions maintained thermal contact between the Mn and the cell walls. The populations of the various trapped states were monitored by absorption spectroscopy on the $^6S_{5/2} - {}^6P_{7/2}$ transition (403 nm).

Fig. 1(c) shows a typical spectrum of trapped Mn. The data is fit (solid line) by calculating the absorption for transitions between the $^6S_{5/2}$ ground state manifold and the $^6P_{7/2}$ excited state manifold as described elsewhere.[17] The fit only includes transitions from ground states $|1\rangle$ through $|6\rangle$ i.e., states which in the large *B* limit correspond to the maximally trapped $m_S = 5/2$ states. No spectroscopic evidence of other states (i.e. $|7\rangle$ through $|36\rangle$) was seen, presumably due to the weaker trapping potential and more rapid spin-exchange for these states. The slight discrepancy in the peak near zero detuning (Fig. 1(c)) is likely due to slight misalignment of the probe laser with the trap.

Fig. 2 shows the time dependence of the populations of the six trapped states $|1\rangle$ - $|6\rangle$. The population of each state is extracted from spectra and fits like the ones in Fig. 1(c). The data in Fig. 2(a) were taken at $T = 855$ mK and trap depth $B_{trap} = 3.9$ T, while the data in Fig. 2(b) were taken at $T = 480$ mK and $B_{trap} = 2.0$ T. These parameters were chosen to give identical values of $\eta \equiv g m_S \mu_B B_{trap} / k_B T$ for the two data sets. This ensures that atom loss over the top of the trap is the same in both data sets and that differences in atom loss reflect the $B$- and $T$- dependence of the atoms' collisional properties. Fits to the spectra showed that the atoms remained in good thermal contact with the cell walls, as mentioned above.

It is clear that the decay does not have the single-exponential form expected for loss due to collisions with background gas. This is consistent with the large mass ratio between $^{55}$Mn and $^3$He and the fact that the trap is much deeper than the $^3$He atoms' kinetic energy.

We model the time-dependence of the six trapped states' populations with six coupled rate equations which include all possible two-body inelastic processes:

$$\dot{n}_1 = \sum_{\{x,y\}} \Gamma^{(1)}_{x,y} n_x n_y \;\; ; \;\; \dot{n}_2 = \sum_{\{x,y\}} \Gamma^{(2)}_{x,y} n_x n_y \;\; ; \;\; \ldots \;\; ; \;\; \dot{n}_6 = \sum_{\{x,y\}} \Gamma^{(6)}_{x,y} n_x n_y \qquad (1)$$

Here $n_z(t)$ is the population of the state $|z\rangle$ and $\Gamma^{(z)}_{x,y}$ is the rate coefficient corresponding to changes in the population of state $|z\rangle$ due to collisions between atoms in states $|x\rangle$ and $|y\rangle$. The indices $x$, $y$, and $z$ each run from 1 to 6, as these are the only states present in the trap. The sums are over the 21 distinguishable pairs of x and y. This leads to 126 independent rate coefficients $\Gamma^{(z)}_{x,y}$.

We calculate the rate coefficients $\Gamma^{(z)}_{x,y}$ in the Born approximation:[18]

$$\Gamma^{(z)}_{x,y} \propto \iint \sum_{\{x',y'\}} |\langle x', y', \mathbf{k}'|V_c|x, y, \mathbf{k}\rangle|^2 \Lambda e^{-\hbar^2 k^2/2mk_B T} \delta(\Delta E) d^3\mathbf{k}' d^3\mathbf{k} \qquad (2)$$

where the constant of proportionality is the same for all x, y, and z and is absorbed into the fitting parameters described below. The sum in Eq. 2 is over the distinguishable pairs of final atomic states x' and y'. The indices x' and y' run from 1 to 36 to account for all possible inelastic processes. $V_c$ is the operator describing the atom-atom interaction, $\Lambda \equiv \delta_{x',z} + \delta_{y',z} - \delta_{x,z} - \delta_{y,z}$ counts the net change in the population of state $|z\rangle$ for each term in the sum, $\mathbf{k}$ and $\mathbf{k}'$ are the atoms' initial and final relative wave vectors ($\mathbf{k}$ is assumed to have a Boltzmann distribution), $\Delta E$ is the energy difference between the initial and final states, and the Dirac-delta function in Eq. (2) ensures energy conservation (it constrains $\mathbf{k}'$ for a given $\mathbf{k}$, x, y, x', y' and B). The atoms' mass is m.

The dominant inelastic processes are expected to be magnetic dipole and spin-exchange, so we take $V_c = V_d + V_e$. The magnetic dipole interaction is $V_d = \mu_0 \mu_B^2 g^2 (\mathbf{S}_1 \cdot \mathbf{S}_2 - 3(\mathbf{S}_1 \cdot \hat{\mathbf{r}})(\mathbf{S}_2 \cdot \hat{\mathbf{r}}))/4\pi r^3$ and the spin-exchange interaction is $V_e = J(r)\mathbf{S}_1 \cdot \mathbf{S}_2$ where $\mathbf{r}$ is the interatomic separation, $r = |\mathbf{r}|$, $\hat{\mathbf{r}} = \frac{\mathbf{r}}{r}$, and $J(r)$ is a function which depends upon the overlap of the two atoms' valence electron wavefunctions, but is not known accurately for Mn dimers.

Because the overwhelming majority of the atoms are in large $B$ fields where $\varepsilon \equiv a/g\mu_B B \ll 1$ ($a$ is the Mn magnetic dipole hyperfine coupling), we use the $|m_S, m_I\rangle$ basis for single-atom states. In the limit $\varepsilon \to 0$, $V_e$ generates no inelastic transitions and $V_d$ only generates transitions in which $m_I$ is unchanged and one or both of the atoms' $m_S$ is raised or lowered by unity. In this case each of the rate equations (Eq. 1) simplifies to $\dot{n}_z = \Gamma_d n_z n_{tot}$ where $n_{tot} = \sum_{z=1}^{6} n_z$ and $\Gamma_d$ is a constant. This system of equations can be solved to give $n_z(t) = n_z(0)/(1 + n_z(0)\Gamma_d t)$, i.e., all six populations having the same time dependence. It is clear from Fig. 2 that this is not the case.

To understand the different time dependences of the trapped states, we evaluate the matrix elements $\langle x', y', \mathbf{k}'|V_c|x, y, \mathbf{k}\rangle$ appearing in Eq. 2, assuming the collisions take place in a region of the trap where $\varepsilon \ll 1$. We calculate the spin part of the single-atom wavefunctions to first order in $\varepsilon$. We ignore symmetrization of the two-atom states because of the large number of partial waves and hyperfine states in the trap.

It is now straightforward to contract $V_c$ between the initial and final spin states and to perform the sum over final spin states in Eq. 2. The final step in evaluating the rate coefficients $\Gamma_{x,y}^{(z)}$ is to contract $V_c$ between the initial and final momentum states and integrate the result over $\mathbf{k}$ and $\mathbf{k}'$. The five integrals involving $V_d$ (each describes a different type of transition) could not be performed analytically. As a first approximation, we set them all equal to a parameter ($A_d$) which we take as a fitting parameter. This has the advantage of implicitly including atom-atom interactions which have the same spin dependence as dipole-dipole but different spatial dependence (e.g., second-order spin-orbit coupling).

Since $J(r)$ is unknown, the integral involving $V_e$ is taken as another fitting parameter ($A_e$). This gives us expressions for all 126 rate coefficients involving $A_d$, $A_e$ and $\varepsilon$. The approach described here amounts to using the Born approximation to determine (to order $\varepsilon^2$) the *fractional* change in the dipolar and spin-exchange relaxation rate coefficients due to hyperfine-induced mixing with $m_S \neq \frac{5}{2}$ states while keeping the *overall* rates as fitting parameters.

The atom density varies throughout the trap, as do the rate coefficients. The rate coefficients' spatial dependence enters through $\varepsilon$, which depends upon $B$ and hence the position in the trap. Averaging the rate coefficients over the atoms' Boltzmann distribution gives the usual factor of $\frac{1}{8}$ for the $B$-independent term (i.e., the hyperfine-unmodified dipolar contribution), while the $B$-dependent ($\propto \varepsilon^2$) terms lead to an "average" magnetic field $B_{avg} = 2k_BT/g\mu_B m_S$ or an "average" $\varepsilon_{avg} = am_S/2k_BT$.

Before attempting to fit the data quantitatively, we note that in Fig. 2 the decay shows a stronger $m_I$-dependence at lower $T$ (and hence $B$). This indicates that the dominant angular momentum changing collisions are sensitive to the residual hyperfine interactions, as we would expect for spin-exchange. To confirm this we first fit the data allowing only $A_d$ to vary and setting $A_e = 0$. These fits failed to reproduce the basic trends in the data, providing qualitative evidence that both hyperfine-modified dipolar and spin-exchange collisions are relevant.

Next we fixed $\varepsilon = \varepsilon_{avg}$ and used $A_d$ and $A_e$ as fit parameters. This reproduces the data fairly well, capturing the important trends in Fig. 2, such as states with lower $m_I$ decaying more quickly (relative to their initial population) and the fact that this behavior is more pronounced at lower $T$ (corresponding to lower $B_{avg}$ hence higher $\varepsilon_{avg}$ and more hyperfine-induced mixing). The fitting is simplified because in practice $A_d$ sets the overall loss rate and $A_e$ sets the differences among the populations' loss rates.

A weakness of this model is that it is valid only for small $\varepsilon$, i.e. for $B$ much larger than the hyperfine splitting. This condition is satisfied in the overwhelming majority of the trap *volume*, but the atom density and hyperfine-induced mixing are both highest at the trap center where $\varepsilon$ is no longer small and the single-atom spin wavefunctions used above are not accurate. Thus our model does not treat collisions near the trap center correctly and so the trap loss may be dominated by a different value of $\varepsilon$ than $\varepsilon_{avg}$. To account for this we also fit the data treating $\varepsilon$ as a fitting parameter. In practice this allows for a small amount of additional fine-tuning of the $m_I$-dependence of the loss rates. The result is shown as the solid lines in Fig. 2, and reproduces the data quite accurately. The fit in Fig. 2(a) gives $A_d = 4.2 \times 10^{-8}$ cm$^{3/2}$/s$^{1/2}$, $A_e = 3.5 \times 10^{-7}$ cm$^{3/2}$/s$^{1/2}$, and $\varepsilon = 0.0090$. By comparison the *ab initio* value $\varepsilon_{eff} = 0.0047$. The fit in Fig. 2(b) gives $A_d = 2.6 \times 10^{-8}$ cm$^{3/2}$/s$^{1/2}$, $A_e = 1.5 \times 10^{-7}$ cm$^{3/2}$/s$^{1/2}$, and $\varepsilon = 0.028$ (compared with $\varepsilon_{eff} = 0.0085$).

Treating $\varepsilon$ as a fitting parameter is an *ad hoc* solution to the breakdown of perturbation theory at the trap center. However, the fits in Fig. 2 show an improved agreement with the data and imply a value for the effective $\varepsilon$ slightly *greater* than $\varepsilon_{eff}$. Both of these features are consistent with collisions near the center of the trap playing an enhanced role in the trap loss. A more complete treatment of collisions at large $\varepsilon$ would need to consider the full Zeeman structure of the ground state manifold (Fig. 1(a)), a task beyond the scope of this paper.

We note as an aside that one-body losses should reflect the presence of the many level crossings at low *B*. In particular, crossings between states from the $m_S$ = 5/2 and $m_S$ = 3/2 manifolds will lead to Majorana-type loss from spherical surfaces in addition to the usual point at the trap center. One example is the crossing of $|5\rangle$ and $|7\rangle$ at *B* = 0.02 T (Fig. 1(a)). Interestingly, $|1\rangle$ (which asymptotes to $|m_S = \frac{5}{2}, m_I = \frac{-5}{2}\rangle$ as $\varepsilon \to 0$) has no level crossings, even at *B* = 0 where it becomes the *F* = 0 singlet. As a result, this state does not undergo any Majorana losses, even in an anti-Helmholtz trap. However it is susceptible to spin-exchange and dipolar losses.

Comparing the results of this analysis with results for non-submerged-shell atoms is not entirely trivial, as the fits in Fig. 2 yield 126 different rate coefficients. In addition, our measurements were made in the small-$\varepsilon$ regime, the opposite from most atom-trapping experiments. To see whether the spin-exchange rate coefficients inferred from Fig. 2 are influenced by the submerged shell nature of Mn, we must compare them with spin-exchange rate coefficients for non-submerged-shell atoms at comparable values of $\varepsilon$. This is possible for $^{23}$Na, where the rate coefficient $\Gamma^{(7),Na}_{6,7,e}$ for spin-exchange collisions between Na atoms in states $|6\rangle_{Na}$ and $|7\rangle_{Na}$ (which asymptote to $|m_S = \frac{1}{2}, m_I = \frac{1}{2}\rangle_{Na}$ and $|m_S = \frac{1}{2}, m_I = -\frac{1}{2}\rangle_{Na}$) have been calculated as a function of *B*.[19] Extrapolating the results of Ref. 19 gives $\Gamma^{(7),Na}_{6,7,e} = 4 \times 10^{-15}$ cm$^3$/s for $\varepsilon$ = 0.009, equivalent to Fig. 2a, and $\Gamma^{(7),Na}_{6,7,e} \approx 4 \times 10^{-14}$ cm$^3$/s for $\varepsilon$ = 0.028, equivalent to Fig. 2b.

We take the Mn states $|5\rangle$ and $|4\rangle$ as the equivalent states - i.e., those with one and two lower $m_I$ than the stretched state (Fig. 1(a)). The component of the rate coefficient $\Gamma^{(5)}_{5,4}$ which is due only to spin-exchange is $\Gamma^{(5)}_{5,4,e} = 1250\varepsilon^2 A_e^2 = 1.2\times 10^{-14}$ cm$^3$/s ($\varepsilon = 0.009$) and $2.2\times 10^{-14}$ cm$^3$/s ($\varepsilon = 0.028$). Thus the rate coefficients for Mn are comparable to Na. Calculations of spin-exchange in atomic H also give rate coefficients comparable to those measured here for Mn.[20] This suggests that the submerged valence of Mn does not strongly suppress spin-exchange, though we note that the values for Na were calculated at $T = 0$ and for only a single partial wave, whereas the data in Fig. 2 were taken at temperatures where several partial waves contribute to the collisions.

The fact that spin-exchange collisions are not suppressed in Mn is somewhat surprising. Spin-exchange arises from the overlap of colliding atoms' valence electronic wavefunctions, and so would seem susceptible to shielding by filled outer orbitals. Whether the results of this work are general or merely peculiar to Mn could be tested by repeating the measurements and analysis presented here for other $L = 0$, $I \neq 0$ atoms and comparing the results for submerged shell species ($^{185}$Re, $^{187}$Re, $^{151}$Eu, and $^{153}$Eu) with non-submerged species (such as alkalis, coinage metals, $^{53}$Cr, $^{95}$Mo, and $^{97}$Mo). It may be that rare-earth atoms whose 4f valences are shielded by both 5s and 6s orbitals achieve stronger spin-exchange suppression than is observed here for Mn.

The fits in Fig. 2 also give the dipolar rate coefficients. As with spin-exchange, the hyperfine-induced mixing leads to many different rate coefficients, but the $\varepsilon = 0$ value ($\Gamma_d$) is given by $(675/4) A_d^2 = 3.0\times 10^{-13}$ cm$^3$/s (for $\varepsilon = 0.009$) and $= 1.1\times 10^{-13}$ cm$^3$/s (for $\varepsilon = 0.028$). These values are slightly smaller than measured for $^{52}$Cr under similar circumstances,[12] consistent with the slightly smaller magnetic dipole moment of Mn relative to Cr.

In conclusion, we have measured the rate coefficients for spin-exchange and dipolar collisions in the most-low-field-seeking manifold of $^{55}$Mn. The dipolar rate coefficients are comparable to those measured for Cr at similar magnetic fields and temperatures, as we would expect. The spin-exchange rate coefficients are comparable to those calculated for Na and H in Refs. 19 and 20, and do not appear to be suppressed by the submerged valence of Mn.

We gratefully acknowledge helpful discussions with Roman Krems, Jonathan Weinstein, and Robert Michniak. This work was carried out at Harvard University as part of the NSF Center for Ultracold Atoms.

**Figure Captions**

**Figure 1** (two columns): (a) The ground state manifold of $^{55}$Mn. The Zeeman splitting of the 36 levels in the ground state of Mn is shown, along with the quantum numbers in the small- and large-$B$ limits. (b) Schematic of the experimental cell. The cell is thermally linked to a $^3$He refrigerator. For each measurement the lower chamber is filled with $^3$He from the reservoir and solid Mn is ablated. Then the valve is opened and the $^3$He is removed by the sorption pump. (c) Spectrum of trapped Mn. The solid line is a fit which includes the Maxwell-Boltzmann distribution of the atoms in the magnetic trap (calculated numerically from the coil geometry), the finite radius of the probe beam, its small offset from the trap center, the orientation of its polarization, the atoms' Doppler and lifetime broadening, and the Clebsch-Gordan coefficients of the optical transitions.

**Figure 2**: (single column, color): Trap loss of the various hyperfine states. (a) Peak density $n(t)$ of each of the six hyperfine states in the trap versus time at $T = 855$ mK and $B_{trap} = 3.9$ T. The lines are fits described in the text. (b) same as (a) but for $T = 480$ mK and $B_{trap} = 2.0$ T.

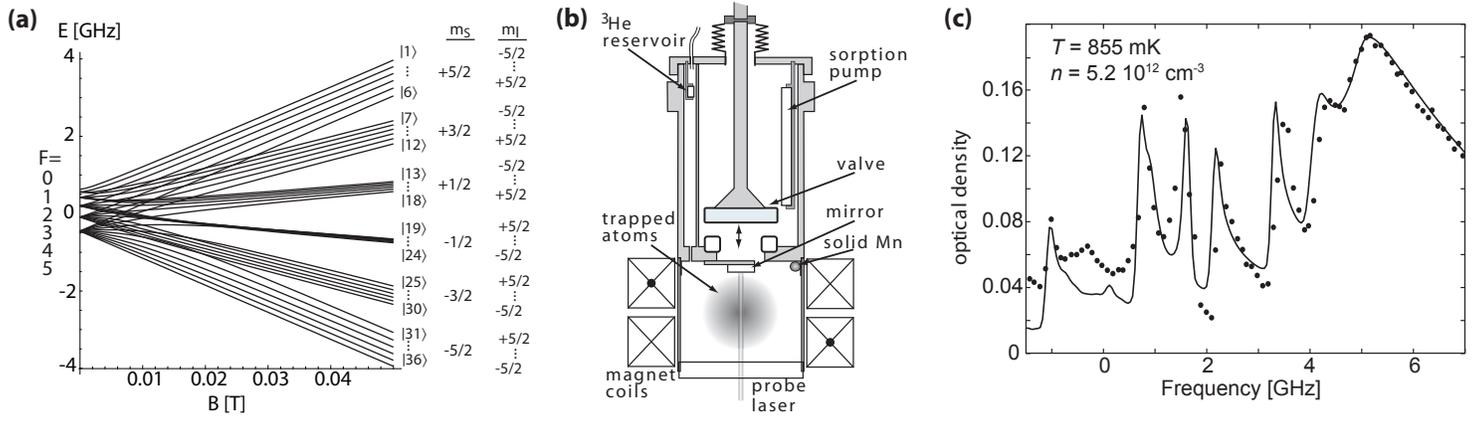

**Figure 1 (two columns)**

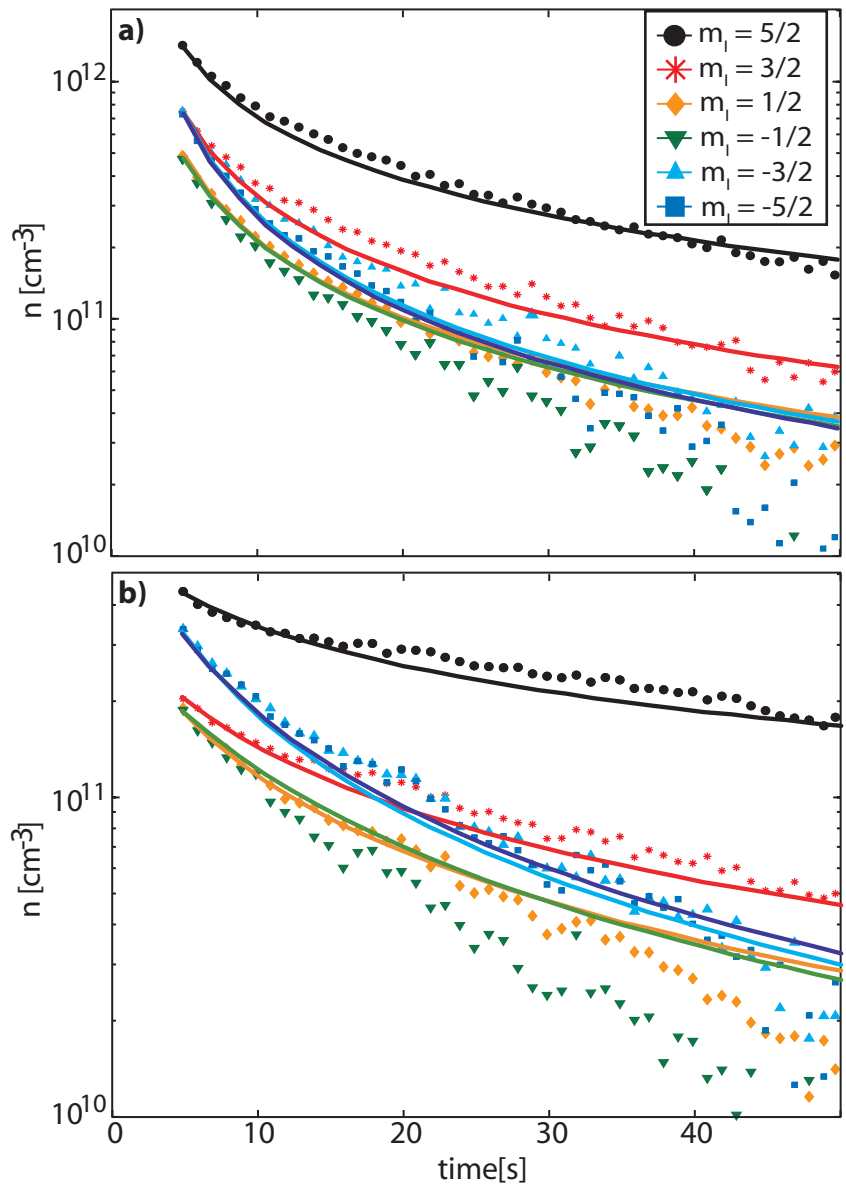

**Figure 2 (color)**